\documentclass[aps,prb,showpacs,two column,showkeys,superscriptaddress,preprintnumbers]{revtex4}
\usepackage{amsmath}
\usepackage{txfonts}
\usepackage[titletoc]{appendix}
\usepackage{amssymb}
\usepackage{array}
\usepackage{mathrsfs}
\usepackage{subfigure,overpic}
\usepackage{dcolumn}
\usepackage{multirow}
\usepackage{array}
\usepackage{booktabs}
\usepackage{setspace}
\usepackage{diagbox}
\usepackage{float}
\usepackage{threeparttable}
\usepackage{color}

\begin{document}

\title{The prolonged decay of RKKY interactions by interplay of relativistic and
non-relativistic electrons in semi-Dirac semimetals}
\author{Hou-Jian Duan}
\affiliation{Guangdong Provincial Key Laboratory of Quantum Engineering and Quantum
Materials, School of Physics and Telecommunication Engineering, South China
Normal University, Guangzhou 510006, China}
\affiliation{Guangdong-Hong Kong Joint Laboratory of Quantum Matter, Frontier Research
Institute for Physics, South China Normal University, Guangzhou 510006, China}
\author{Yan-Yan Yang}
\affiliation{Guangdong Provincial Key Laboratory of Quantum Engineering and Quantum
Materials, School of Physics and Telecommunication Engineering, South China
Normal University, Guangzhou 510006, China}
\author{Shi-Han Zheng}
\affiliation{College of Automation, Zhongkai University of Agriculture and Engineering,
Guangzhou 510225, China}
\author{Chang-Yong Zhu}
\affiliation{Guangdong Provincial Key Laboratory of Quantum Engineering and Quantum
Materials, School of Physics and Telecommunication Engineering, South China
Normal University, Guangzhou 510006, China}
\affiliation{School of Intelligent Engineering, Shaoguan University, Shaoguan 512005,
China}
\author{Ming-Xun Deng}
\author{Mou Yang}
\author{Rui-Qiang Wang}
\email{wangruiqiang@m.scnu.edu.cn}
\affiliation{Guangdong Provincial Key Laboratory of Quantum Engineering and Quantum
Materials, School of Physics and Telecommunication Engineering, South China
Normal University, Guangzhou 510006, China}
\affiliation{Guangdong-Hong Kong Joint Laboratory of Quantum Matter, Frontier Research
Institute for Physics, South China Normal University, Guangzhou 510006, China}

\begin{abstract}
The Ruderman-Kittel-Kasuya-Yosida (RKKY) interaction has been extensively explored in isotropic Dirac systems with linear dispersion, which typically follows an exponent decaying rate with the impurity distance $R$, i.e., $J\propto 1/R^d$ ($1/R^{2d-1}$) in $d$-dimensional systems at finite (zero) Fermi energy. This fast decay makes it rather difficult to be detected and limits its application in spintronics. Here, we theoretically investigate the influence of anisotropic
dispersion on the RKKY interaction, and find that the introduction of non-relativistic dispersion in semi-Dirac semimetals (S-DSMs) can significantly prolong the decay of the RKKY interaction and can remarkably enhance the Dzyaloshinskii-Moriya interaction around the relativistic direction. The underlying physics is attributed to the highly increased density of states in the linear-momentum direction as a result of the interplay of relativistic and non-relativistic electrons. Furthermore, we propose a general formula to determine the decaying rate of the RKKY interaction, extending the typical formula for isotropic DSMs. Our results suggest that the S-DSM materials are a powerful platform to detect and control the magnetic exchange interaction, superior to extensively adopted isotropic Dirac systems.

\end{abstract}

\maketitle

%%\altaffiliation{Electronic address: yang.mou@hotmail.com}

%%%%%%%%%%%%%%%%%%%%%%%%%%%%%%%%%%%%%
Over the past decades, the Ruderman-Kittel-Kasuya-Yosida (RKKY) interaction
has been extensively studied in a variety of materials, e.g., graphene\cite%
{graphene1,graphene2,graphene3,graphene4,graphene5,graphene6}, $\alpha $-$\mathcal{T}_{3}$ model\cite{alpha}, Dirac/Weyl semimetals (DSMs/WSMs)\cite{Weyl1,Weyl2,Dirac0}, phosphorene\cite{phosphorene2}, edge/surface bands of
topological materials\cite{topo2,Dirac1,surface1,surface2,surface3,surface4} and so on.
These researches have showed a potential application for the RKKY
interaction to realize magnetization in non-magnetic materials. A typical
example is the realization of quantized anomalous Hall effect in topological
insulators\cite{anous,topo2}. Moreover, the RKKY interaction has been proven
to characterize the intrinsic properties of materials, e.g., the band
topology\cite{surface1,NLSM}, rich spin textures\cite{Weyl1}, and the Rashba
splitting\cite{Shiranzaei}. Although the RKKY interaction has wide prospect
in the area of spintronics, there are still many obstacles to overcome. One
is that the RKKY interaction is too weak to be detected since it usually
decays fast with the increased impurity distance $R$. For example, in
materials with isotropic linear dispersion, the RKKY interaction presents a
fast decaying law as $\mathrm{sin}\left( 2k_{F}R\right) /R^{d}$ ($1/R^{2d-1}$%
) at finite (zero) Fermi energy. To overcome this obstacle, new research
perspectives are expected. In addition to some special devices (e.g., the PN
junction\cite{pn}), materials with peculiar dispersions are promising
candidates to realize the prolonged RKKY interaction.
\par
Recently, semi-Dirac semimetals (S-DSMs) have attracted more and more
attention in condensed matter physics due to their highly anisotropic
electronic structure. Different from the isotropic linear dispersions around
the Dirac points, the low-energy model of S-DSM exhibits a linear dispersion
in some directions but disperses quadratically in the others, which allows
the coexistence of relativistic and non-relativistic electrons. This
peculiar dispersion leads to many new physical properties, such as the
anisotropic transports\cite{Adroguer,SPark}, nonsaturating large
magnetoresistance\cite{Niu}, unique optical properties\cite%
{Mawrie1,Carbotte,Dai}, and quantum thermoelectrics\cite{Mawrie2}.
Nevertheless, the magnetic property, especially the RKKY interaction between
magnetic impurities, with respect to S-DSMs receives no attention. It is
expected that the strong anisotropic dispersion of the S-DSMs will affect
the RKKY interaction significantly.
\par
In this Letter, we theoretically investigate the influence of anisotropic
dispersion on the RKKY interaction, and take S-DSMs as examples of
anisotropic structure and compare them with isotropic DSMs/WSMs. We find
that the decaying rate of the RKKY interaction in S-DSMs, including all
types of two or three dimensions, can be highly prolonged along the
relativistic axis, in contrast to the fast-decaying RKKY interaction in
DSMs/WSMs. Furthermore, we propose a general formula to determine the
decaying rate of the RKKY interaction for anisotropic DSMs with arbitrary dimensions.
\par
\emph{RKKY theory-} RKKY interaction describes an indirect exchange
interaction, mediated by itinerant electrons, between two impurities
embedded in the material. We start from the Hamiltonian%
\begin{equation}
H=H_{0}+\lambda \sum_{i=1,2}\mathrm{\mathbf{S}}_{i}\cdot \mathrm{\mathbf{s}}%
_{i},
\end{equation}%
where $H_{0}$ stands for the host materials and $\lambda $ is spin exchange
between the itinerant electron spin $\mathrm{\mathbf{s}}_{i}$ and the
impurity spin $\mathrm{\mathbf{S}}_{i}$, located at $\mathbf{r}_{i}$. For
weak coupling $\lambda $, we can use the standard perturbation theory\cite%
{rkky1,rkky2,rkky3}, and up to the second order of $\lambda $, the
RKKY interaction at zero temperature can be calculated by,
\begin{equation}
H_{RKKY}=-\frac{\lambda ^{2}}{\pi }\mathrm{Im}\int_{-\infty }^{u_{F}}\mathrm{%
Tr}[\left( \mathrm{\mathbf{S}_{1}}\cdot \mathbf{\sigma }\right) G\left(
\mathbf{R},\omega \right) \left( \mathrm{\mathbf{S}_{2}}\cdot \mathbf{\sigma
}\right) G\left( -\mathbf{R},\omega \right) ]d\omega ,
\end{equation}%
where $G\left( \pm \mathbf{R},\omega \right) $ is the retarded Green's
function with respect to $H_{0}$ in real space and $u_{F}$ is the Fermi
energy. After tracing the spin degrees of freedom in Eq. (2), the RKKY
interaction can be written in the form of
\begin{equation}
H_{RKKY}=\sum_{\alpha ,\beta =x,y,z}J^{\alpha \beta }{S_{1}^{\alpha }}{%
S_{2}^{\beta }}
\end{equation}%
with
\begin{equation}
J^{\alpha \beta }=-\frac{\lambda ^{2}}{\pi }\mathrm{Im}\int_{-\infty
}^{u_{F}}\sum_{i,j=0,x,y,z}\Lambda _{ij}\left( \mathbf{R},\omega \right)
\mathrm{Tr[}\sigma _{\alpha }\sigma _{i}\sigma _{j}\sigma _{\beta }]d\omega ,
\end{equation}%
where we denote matrix of Green's function as $G\left( \pm \mathbf{R},\omega
\right) =\sum_{i=0,x,y,z}G_{i}\left( \pm \mathbf{R},\omega \right) \sigma
_{i}$ and $\Lambda _{ij}\left( \mathbf{R},\omega \right) =G_{i}\left(
\mathbf{R},\omega \right) G_{j}\left( -\mathbf{R},\omega \right) $. In Eq.
(4), the trace part determines the spin exchange type and $\Lambda
_{ij}\left( \mathbf{R},\omega \right) $ determines the decaying rate with
impurity distance $\mathbf{R}=\mathbf{r_{2}}-\mathbf{r}_{1}$. Physically, $%
\Lambda _{ij}\left( \mathbf{R},\omega \right) $ characterizes the spin $%
\sigma _{i}$ disturbance\cite{density} of site $\mathbf{r}_{1}$ as a
response to the spin $\sigma _{j}$ of site $\mathbf{r}_{2}$.
\par
\emph{Universal RKKY decay for isotropic materials-} Obviously, the decay of
the RKKY interaction is determined by $\Lambda _{ij}\left( \mathbf{R},\omega
\right) $, which is closely related to material type. Considering a $d$%
-dimensional system with linear Dirac cone, $H_{\mathrm{0}}=v\mathbf{k}\cdot
\mathbf{\sigma }$, where $v$ is the Fermi velocity, we have
\begin{equation}
G_{i}\left( \mathbf{R},\omega \right) =\frac{1}{R^{\left( d-1\right) /2}}%
\omega ^{\left( d-1\right) /2}e^{i\omega R/v},
\end{equation}%
with $R=|\mathbf{R}|$. The resulting RKKY interaction is
\begin{eqnarray}
J^{\alpha \beta }\left( u_{F}\neq 0\right) &\propto &\frac{1}{R^{d}}\mathrm{%
sin}\left( \frac{2u_{F}R}{v}\right) , \\
J^{\alpha \beta }\left( u_{F}=0\right) &\propto &\frac{1}{R^{d+\zeta }}=%
\frac{1}{R^{2d-1}},
\end{eqnarray}%
where $\zeta=d-1$ is the exponent of the density of states (DOS) of DSMs.
The above laws have been extensively reported in isotropic systems with
linear dispersion, such as graphene\cite%
{graphene1,graphene2,graphene3,graphene4,graphene5,graphene6}, and DSMs/WSMs\cite{Weyl1,Weyl2,Dirac0}. If the isotropic system is usual
quadratical dispersion $H_{0}=v' k^2\sigma _{0}$, the RKKY interaction
follows a law\cite{eletrongas1,eletrongas2} as $J^{\alpha \beta }\left(
u_{F}=0\right) =0$ and $J^{\alpha \beta }\left( u_{F}\neq 0\right) \propto
\mathrm{sin}\left( 2R\sqrt{u_{F}/v'}\right) /R^{d}$ which shares a
same decaying rate but with a different oscillation as compared to Eq. (6).
\par
\emph{Universal RKKY decay for S-DSM materials-} The above argument is valid
only for isotropic systems since $e^{i\omega R/v}$ in Eq. (5) is obtained
under the $\theta _{R}$-independent condition [$\mathrm{tan}\left( \theta
_{R}\right) =R_{y}/R_{x}$]. In S-DSMs with anisotropic dispersions, the
corresponding Green's function is direction-dependent.
\par
We employ a general Hamiltonian of a minimal low-energy model of S-DSM
\begin{equation}
H_{\mathrm{S-DSM}}=v_{p}\left( k_{x}^{2}+\xi _{1}k_{y}^{2}\right) \sigma
_{z}+v_{l}\left(\xi _{2}k_{y}\sigma _{x}+k_{z}\sigma _{y}\right),
\end{equation}%
which collects all S-DSM models in 2D and 3D materials: (1) Case I: for $\xi
_{1}=0$ and $\xi _{2}=0,$ it reduces to S-DSM in $2D$ case\cite%
{Victor1,Victor2,Banerjee,Banerjee2,Qiuzi,Zhai}. (2) Case II: for $\xi _{1}=0$ and $\xi _{2}=1$,
it is linear momentum along two directions and square along the other one,
called as $3D$-type I. (3) Case III: For $\xi _{1}=1$ and $\xi _{2}=0,$ it
is linear momentum along one direction and square along other two
directions, called as $3D$-type II. Noting that the Pauli matrices $\sigma_{x,y,z}$ in Eq. (8) can act either in pseudo-spin basis\cite{Victor1,Victor2,Banerjee} or real-spin basis\cite{Banerjee2,Qiuzi,Zhai}, depending on specific materials. For example, the pseudo-spin 2D S-DSM can be extracted from $\left({\rm TiO_2}\right)_5/\left({\rm VO_2}\right)_3$ multilayer system\cite%
{Victor1,Victor2,Banerjee}, and the
real-spin one can be obtained from the topological surface band under the effect of a helical spin density wave\cite{Qiuzi} or a spiral magnetization superlattice\cite{Zhai}. Without loss of generality, we firstly assume that the Hamiltonian of Eq. (8) is written in real-spin basis. Later in this paper, the effect of the pseudo-spin case on the RKKY interaction would be discussed.
\par
Different from the numerous candidates for 2D S-DSM, few literatures focus on 3D S-DSMs. The two types of 3D S-DSMs used in our paper can be obtained by applying linearly polarized light in WSMs\cite{photoWSM1} and nodal-line
semimetals\cite{WeiChen} (NLSMs), respectively. For example, one can consider a low-energy model of WSM with broken time-reversal symmetry as\cite{photoWSM1},
\begin{equation*}
H_{\mathrm{WSM}}=v_{p}\left( k_{x}^{2}-m^{2}\right) \sigma
_{z}+v_{l}\left( k_{y}\sigma _{x}+k_{z}\sigma _{y}\right) ,
\end{equation*}%
where $\sigma _{x,y,z}$ refer to the Pauli matrices of the spin degrees of freedom, and two Weyl points are located at $(\pm m,0,0)$. After
introducing a beam of linearly polarized light of frequency $\omega $, a vector potential $A=A(1,0,0){\rm cos}\left( \omega t\right) $ with period $T=2\pi /\omega$ is generated. By applying the Peierls substitution $\hbar k$$\rightarrow $$\hbar k+eA$, the
system Hamiltonian becomes time-dependent. Using the Floquet theory\cite{floquet3} with the off-resonant condition of $A^{2}/\omega \gg 1$, the modified part of the Hamiltonian induced by
light reads as $V_{0}+\sum_{n\geq 1}\left[ V_{+n},V_{-n}\right] /\hbar
\omega +O\left( 1/\omega ^{2}\right) $ with $V_{n}=\frac{1}{T}%
\int_{0}^{T}H(t)e^{-in\hbar \omega t}$, and the effective
Hamiltonian can be written as
\begin{equation*}
H_{\mathrm{WSM}}^{\prime }=H_{\mathrm{WSM}}+\frac{v_{p}e^{2}A^{2}}{2}%
\sigma _{z},
\end{equation*}%
where the term related to $A$ refers to the photoinduced
modification. By setting proper amplitude $A$ of the vector
potential with $e^{2}A^{2}/2=m^{2}$, the Weyl partners are merged
into a point, i.e., the WSM is changed to be the type I of 3D S-DSM [$\xi_1=0$ and $\xi_2=1$ in Eq. (8)]. Similarly, the type II of 3D S-DSM can also be obtained when the
nodal ring of NLSM is shrunk into a point by the linearly polarized light.
\par
The Green's function with respect to $H_{%
\mathrm{S-DSM}}$ of Eq. (8) reads as,
\begin{eqnarray}
G\left(\pm\mathbf{{R},\omega}\right)=\frac{1}{(2\pi)^d}\int d^d{\mathbf{k}}%
\frac{\omega\sigma_0+H_{\mathrm{S-DSM}}}{\left(\omega+i0^+\right)^2-E_{%
\mathrm{S-DSM}}^2}e^{\pm i\mathbf{k}\mathbf{R}},
\end{eqnarray}
with $E_{\mathrm{S-DSM}}$ denoting the eigenenergy of the Hamiltonian $H_{%
\mathrm{S-DSM}}$. The analytical results of $G_{i}\left( \pm \mathbf{{R}%
,\omega }\right) $ for different S-DSMs can be obtained after some algebraic
calculations (see the supplemental material\cite{appendix}).
\par
\begin{table}[th]
\caption{The RKKY components $J_{l}^{\protect\alpha%
\protect\beta}$ in S-DMSs with impurities in the relativistic axis, where
coefficients unrelated to the decaying law, and the oscillation $A{\sin}\left(2k_FR\right)+B{\cos}\left(2k_FR\right) $ with $k_F=u_F/v_l $  for the case of $u_F\neq 0$ are dropped. For $u_F\neq 0$, some slowly decaying RKKY components (e.g., $R^{-3/2}$ for 2D S-DSM) vanishes and the higher-order (e.g., $R^{-2}$ for 2D S-DSM)
term plays a leading role, but the slowly decaying law still exists in other RKKY components (see the supplemental material\cite{appendix}). }\centering
\begin{spacing}{1.7}
\begin{tabular}	{c|c|c|c}
\hline\hline
\diagbox{$u_F$}{$J^{\alpha\beta}_l$}{$H$} & $H^{2D}_{\rm S-DSM}$  & $H^{3D,{\rm I}}_{\rm S-DSM}$ & $H^{3D,{\rm II}}_{\rm S-DSM}$ \\ \hline
$u_F\neq0$ &    \;\;$  R^{-3/2}+O(R^{-2})$  \; \;& \;\;  $  R^{-5/2}$  \;\;&\;\;   $ R^{-2}+O(R^{-3})$    \\
\hline
$u_F=0$ &  $R^{-2}$ &  $R^{-4}$ &   $R^{-3}$ \\
\hline \hline
\end{tabular}
\end{spacing}
\end{table}
\par
\emph{A: along relativistic direction- }Firstly,\emph{\ }we perform the
calculations for impurities deposited along the relativistic axis (see the
supplemental material\cite{appendix}), and depict the results in Table I.
For this case, the decaying rate of the interaction ($u_{F}\neq0$) with the
distance $R$ between impurities can be expressed as a general form
\begin{equation}
J_{l}^{\alpha \beta }(u_{F}\neq 0)\propto \left[\frac{1}{R^{d-s/2}}+O\left(\frac{1}{R^{d}}\right)\right]%
\left[A{\sin}\left(2k_FR\right)+B{\cos}\left(2k_FR\right)\right],
\end{equation}%
where $k_F=u_{F}/v_{l}$. Similar oscillation $\mathrm{sin}\left( 2Ru_{F}/v_{l}\right) $ is also
found in DSM-type materials\cite%
{graphene1,graphene2,graphene3,graphene4,graphene5,Weyl1,Weyl2,Dirac0}%
. The index $s$ in the above equation labels the number of dimension of
square momentum, namely, $s=0,1,2$ correspond to the DSMs, 2D DSM (or
3D-type I S-DSM), and 3D-type II S-DSM, respectively. For $u_F\neq 0$, we focus on the slowest-decaying RKKY components ($1/R^{d-s/2}$). Obviously, the
positive number $s$ could reduce the decaying rate as compared with
isotropic systems $J_{l}^{\alpha \beta }\propto 1/R^{d}$ in Eq. (6).
Specifically, for finite Fermi energy, the RKKY component of 2D S-DSM falls
off as $R^{-3/2}$, which decays much more slowly than that of doped phosphorene\cite{phosphorene1,phosphorene2} or 2D DSM where $%
J_{2D}^{\alpha \beta }\propto R^{-2}$ [Eq. (6)]. Compared to the fast
decaying rate of $R^{-3}$ in 3D DSMs/WSMs\cite{Weyl1,Weyl2,Dirac0} with $s=0$%
, the interaction in S-DSM exhibits a slowest decaying rate as $R^{-2}$ for
the type II of 3D S-DSMs with $s=2$ and as $R^{-5/2}$ for the type I of 3D
S-DSMs with $s=1$. This is attributed to the non-relativistic term, which
enters into the anisotropic energy $E_{\mathrm{S-DSM}}$ of Eq. (9) and
competes with the relativistic term to contribute a slowly-decaying rate. This law is unexpected since the slowly-decaying RKKY interaction is usually only realized by the edge/surface state\cite{topo2,Dirac1,surface1,surface2,surface3,surface4}, the strain\cite{phosphorene2}, the PN junction\cite{pn}, etc. So far, few reports have discussed the slowly-decaying (or prolonged) RKKY interaction just mediated by the bulk states, without using any other means.
\par
For zero Fermi energy, we find
\begin{equation}
J_{l}^{\alpha \beta }(u_{F}=0)\propto \frac{1}{R^{d-s/2+\zeta}}=\frac{1}{%
R^{2d-1-s}}.
\end{equation}%
Compared to the case of $u_{F}\neq0$, there are two changes: (1) the spatial
oscillation $\mathrm{sin}\left( 2Ru_{F}/v_{l}\right) $ vanishes; (2) the
decaying rate of the interaction is increased by the exponent $\zeta=d-1-s/2$
of the DOS $|\omega|^\zeta$. Similar effect also have been stated in
DSMs/WSMs\cite{Weyl1,Weyl2,Dirac0} but with $s=0$. Thus, $J_{l}^{\alpha
\beta }(u_{F}=0)$ always decays more slowly than that of DSM-type materials%
\cite{graphene1,graphene2,graphene3,graphene4,graphene5,Weyl1,Weyl2,Dirac0},
where $J^{\alpha\beta}(u_{F}=0)\propto1/R^{2d-1}$ in Eq. (7). Noting that
the relation between the decaying rate and the dimension $s$ of
non-relativistic terms is still similar to the case of $u_{F}\neq0$, i.e,
S-DSMs with larger $s$ would result in a slower decaying rate, as listed in
Table I.
\begin{table}[th]
\caption{The RKKY components $J_{p}^{\protect\alpha
\protect\beta }$ in S-DMSs with impurities in the non-relativistic axis,
where
coefficients unrelated to the decaying law, and the oscillation $C{\sin}\left(2k'_FR\right)+D{\cos}\left(2k'_FR\right) $ with $k'_F=\sqrt{u_F}/\sqrt{v_p}$ for the case of $u_F\neq 0$ are dropped. For $u_F\neq 0$, some slowly decaying RKKY components (e.g., $R^{-2}$ for 2D S-DSM) vanishes and the higher-order
term (e.g., $R^{-3}$ for 2D S-DSM) plays a leading role, but the slowly decaying law still exists in other RKKY components (see the supplemental material\cite{appendix}).}\centering
\begin{spacing}{1.7}
\begin{tabular}	{c|c|c|c}
\hline\hline
\diagbox{$u_F$}{$J_p^{\alpha\beta}$}{$H$}  & $H^{2D}_{\rm S-DSM}$ & $H^{3D,\rm I}_{\rm S-DSM}$ & $H^{3D,\rm II}_{\rm S-DSM}$  \\ \hline
$u_F\neq0$ & \;\;\;$  R^{-2}+O(R^{-3})$ \;\;\;  & \;\;\;  $ R^{-3}$ \;\;\;  &\;\;\; $R^{-3}+O(R^{-4})$ \;\;\;  \\
\hline
$u_F=0$ &  $R^{-4}$&  $ R^{-8}$ & $ R^{-6}$ \\
\hline \hline
\end{tabular}
\end{spacing}
\end{table}
\par
\emph{B: along non-relativistic direction-} Compared to the case with
impurities in the relativistic axis, a faster decaying-rate is exhibited for
the RKKY component $J_{p}^{\alpha \beta }$ when impurities are deposited in
the non-relativistic axis. Performing the similar calculations, we find
\begin{eqnarray}
J_{p}^{\alpha \beta }(u_{F} \neq 0)&\propto& \left[\frac{1}{R^{d}}+O\left(\frac{1}{R^{d-1}}\right)\right]\left[C{\sin}\left(2k'_FR\right)+D{\cos}\left(2k'_FR\right)\right] \\
J_{p}^{\alpha \beta }(u_{F} =0)&\propto& \frac{1}{R^{2\left( 2d-1-s\right) }}
,
\end{eqnarray}%
where $k'_F=\sqrt{u_F}/\sqrt{v_p}$. All the results for different S-DSM models are shown in Table II. For $u_F\neq 0$, we focus on the slowest-decaying RKKY components ($1/R^{d}$). It is found that the decay of the interaction $J_{p}^{\alpha \beta }(u_{F}\neq
0)$ in S-DSMs is only related to the dimensionality $d$, i.e., $R^{-2}$ ($%
R^{-3} $) for 2D (3D) S-DSMs, independent on the dimension $s$ of the
non-relativistic terms. The reason is that, in this impurity configuration,
the interplay of relativistic and non-relativistic electrons is eliminated
by the finite Fermi energy. This would result in the same decaying rate and
the same oscillation $\mathrm{sin}\left( 2R\sqrt{u_{F}}/\sqrt{v_p}\right) $
as that of isotropic systems\cite{eletrongas1,eletrongas2} with quadratical
dispersion. For $u_{F}=0$, all RKKY components of S-DSMs decays faster as
compared to the case of $u_F\neq0$. The reasons are: (1) Similar to $%
J^{\alpha\beta}_l$, the exponent $\zeta=d-1-s/2$ of the DOS $|\omega|^\zeta$
would result in a fast decaying rate for $J^{\alpha\beta}_p$; (2) Compared
to the phase factor $e^{i\omega R/v_{l}}$ of the Green's function $G_l$ [or $%
G_i$ in DSMs of Eq. (5)], the modified phase factor $e^{i\sqrt{\omega}R/%
\sqrt{v_p}}$ of $G_p$ induced by the non-relativistic term would further
accelerate the decaying rate of $J^{\alpha\beta}_p$. As a result, $%
J_{p}^{\alpha \beta }(u_{F}= 0)$ exhibits a fastest decaying rate than that
of $J^{\alpha\beta}_l(u_{F}= 0)$ and $J^{\alpha\beta}(u_{F}= 0)$ in
isotropic systems [Eq. (7)]. Noting that the anisotropic decaying laws shown in Table I and Table II are peculiar as compared to the case of doped phosphorene\cite{phosphorene1}, where the RKKY interaction follows a same decaying law ($R^{-2}$) whether in armchair direction or zigzag direction although the dispersion of the phosphorene is highly anisotropic.
\begin{figure}[th]
\centering \includegraphics[width=0.49\textwidth]{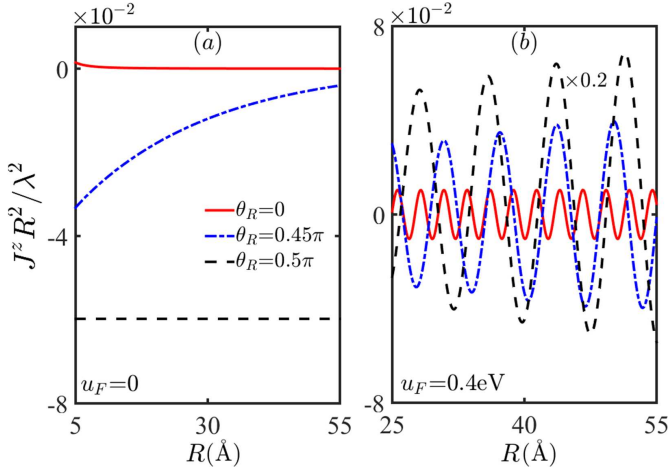}
\caption{(Color online) The $R$-dependent RKKY component $J^z$ with (a) $%
u_F=0$ and (b) $u_F=0.4\mathrm{eV}$ in 2D S-DSMs, where $\protect\theta_R=%
\protect\pi/2$ ($0$) denotes the case with impurities in the relativistic
(non-relativistic) axis. Other parameters\protect\cite%
{Victor1,Victor2,Banerjee} are given by $v_l=0.9873 \mathrm{eV}\mathrm{%
\mathring{A}}$ and $v_p=0.2801 \mathrm{eV}\mathrm{\mathring{A}}^2$.}
\end{figure}
\par
\emph{C: along non-principal directions-}For impurities deposited in
non-principal directions, deviating from relativistic and non-relativistic
axes, the decay of the RKKY interaction can be analyzed only with numerical
calculations. From the above discussions, we know that the decaying rate of
the interaction can be affected heavily by the interplay of relativistic and
non-relativistic electrons, leading to the slower decay in the relativistic
direction than that in the non-relativistic direction. Thus, when impurities
are deposited in a non-principal direction, the intermediate decaying rate
would arise, as shown in Fig. 1, where the interaction decays more fast than
$J_{l}^{\alpha \beta }$ ($\theta_R=\pi/2$) but more slowly than $%
J_{p}^{\alpha \beta }$ ($\theta_R=0$).
\par
\begin{figure}[!ht]
\centering \includegraphics[width=0.42\textwidth]{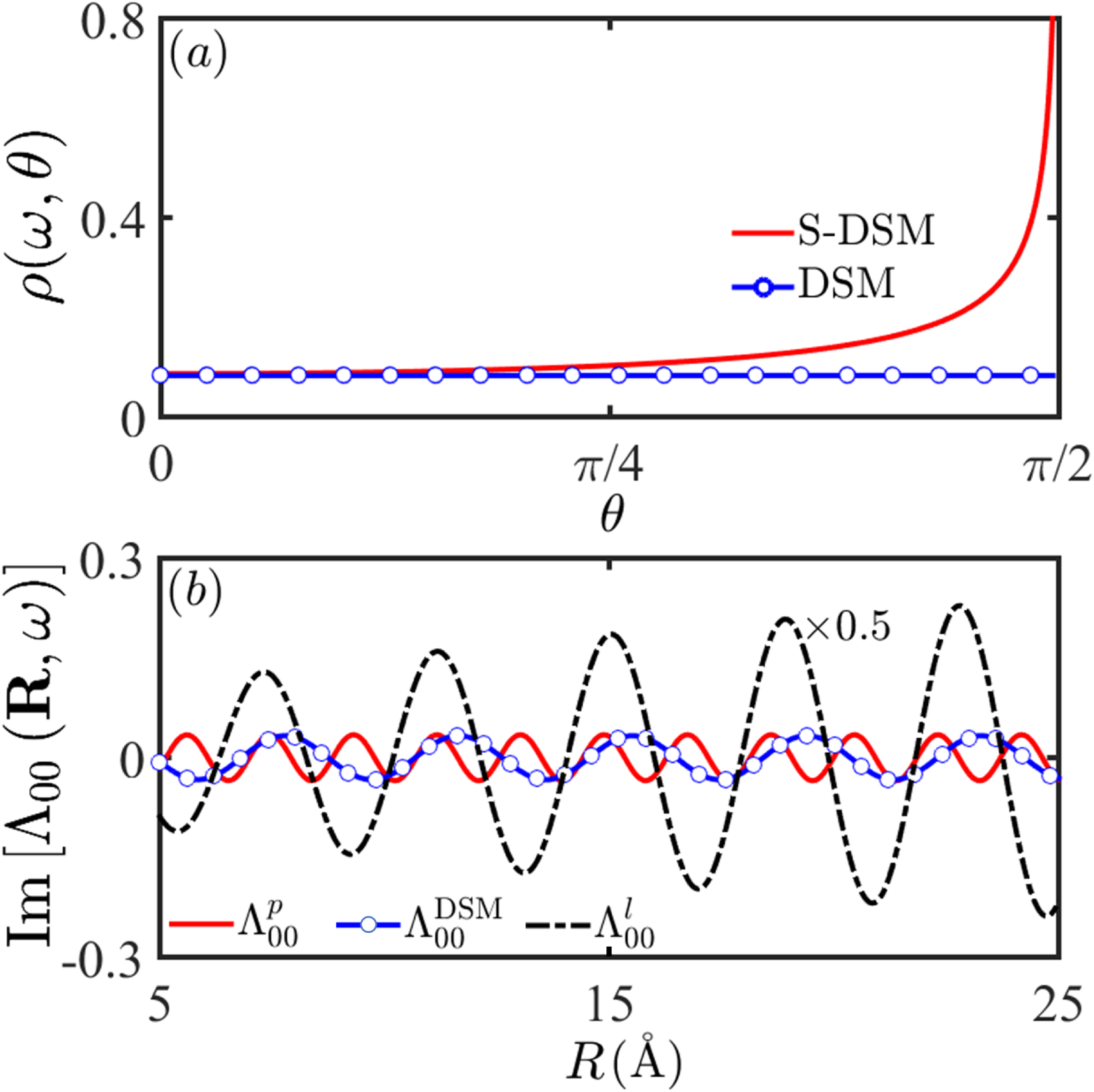}
\caption{(Color online) (a) The direction-dependent DOS $\protect\rho\left(%
\protect\omega,\protect\theta\right)$ with $\mathrm{tan}(\protect\theta%
)=v_lk_z/f_z$ in the Hamiltonian of $f_z\protect\sigma_z+v_lk_z\protect\sigma%
_y$, where $f_z=v_pk_x^2$ ($v_lk_x$) denotes 2D S-DSM (DSM). (b) The
$R$-dependent function ${\rm Im}\left[\Lambda_{00}\left(\mathbf{R},\omega\right)\right]$ with $\omega=0.8\mathrm{eV}$, where
the superscripts $l$ and $p$ denote the cases with impurities in the
relativistic ($\protect\theta_R=\protect\pi/2$) and non-relativistic ($%
\protect\theta_R=0$) axes respectively.}
\end{figure}
\emph{D: Underlying physics- }To more deeply understand the anisotropic RKKY
interaction, we employ the direction-dependent DOS $\rho \left( \omega
,\theta \right) $, which is defined as $\rho \left( \omega ,\theta \right)
=-(1/\pi ){\rm Im}\mathrm{Tr}\int  G\left(\mathbf{k},\omega \right) kdk  $ for 2D case. We obtain $\rho _{_{\mathrm{DSM}}}\left( \omega ,\theta \right) $ $=|\omega |/(\pi^2v_l^2)$ for 2D DSMs and $\rho _{_{\mathrm{S-DSM}}}(\omega ,\theta )=\sqrt{|\omega|}/\left[2\pi^2 v_l\sqrt{v_p{\rm cos}(\theta)}\right]$ for 2D S-DSMs. Obviously, different from the isotropic DSMs, the DOS for S-DSMs is significantly anisotropic, especially along the linear-momentum direction,
i.e., $\theta =\pi /2$, where the DOS of S-DSMs is much larger than that of
DSMs, as shown in Fig. 2(a). The large DOS in the relativistic direction should naturally result in a
slower decaying rate in the real space. To see it, we consider an electron
scattering off a magnetic impurity, whose full electronic Green's function
under Born approximation is modified to be
\begin{equation}
G\left( \mathbf{k,k}^{\prime },\omega \right) =\delta (\mathbf{k-k}^{\prime
})G\left( \mathbf{k},\omega \right) +G\left( \mathbf{k},\omega \right)
\left( \lambda \mathrm{\mathbf{S}_{2}}\cdot \mathbf{\sigma }\right) G\left(
\mathbf{k}^{\prime },\omega \right) .
\end{equation}%
The change of real-space local DOS reads\cite{Mitchell}
\begin{eqnarray}
\delta \rho (\mathbf{R},\omega ) &=&-\frac{\lambda }{\pi }\mathrm{Im}\int d%
\mathbf{q}e^{i\mathbf{q\cdot R}}\sum\limits_{\mathbf{k}}\mathrm{Tr}G\left(
\mathbf{k},\omega \right) \left( \mathrm{\mathbf{S}_{2}}\cdot \mathbf{\sigma
}\right) G\left( \mathbf{k}-\mathbf{q},\omega \right)   \notag \\
&=&-\frac{\lambda }{\pi }\mathrm{ImTr}G\left( \mathbf{R},\omega \right)
\left( \mathrm{\mathbf{S}_{2}}\cdot \mathbf{\sigma }\right) G\left( -\mathbf{%
R},\omega \right) ,
\end{eqnarray}%
in which the decay of $\delta \rho (\mathbf{R},\omega )$ is determined by ${\rm Im}%
[\Lambda _{ij}\left( \mathbf{R},\omega \right)] $. We illustrates the change
of ${\rm Im}\left[\Lambda _{00}\left( \mathbf{R},\omega \right)\right] $ with the impurity
distance $R$ in Fig. 2(b), which shows that the anisotropic
decaying rate in the real space is close related to the anisotropic DOS [Fig. 2(a)].
From above expressions, one also can see that physically, the RKKY
interaction characterizes the change of spin density of itinerant electrons
at site $\mathbf{r}_{1}$ caused by a magnetization at site $\mathbf{r}_{2}$
when electrons complete a round trip from $\mathbf{r}_{1}$ to $%
\mathbf{r}_{2}$, similar to the semiclassical picture of charge density of
isotropic materials\cite{rkky6}. For certain direction, large spin density
will remain more electrons participating the exchange interaction between
magnetic impurities and so prolong the RKKY decay.
\begin{figure}[!ht]
\centering \includegraphics[width=0.42\textwidth]{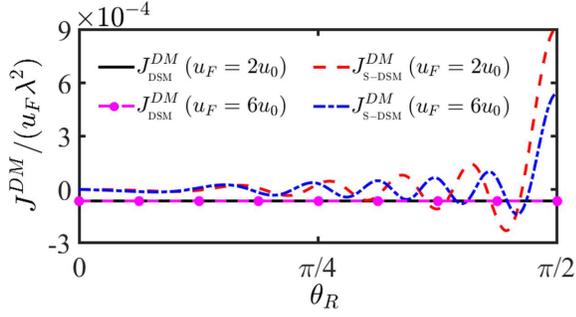}
\caption{(Color online) The direction-dependent DM terms $J^{DM}$ scaled
with $u_F$ in 2D S-DSM and DSM with different $u_F$, where $\protect\theta_R=\protect\pi/2$ ($0$) denotes the relativistic (non-relativistic) axis, and $u_0=\pi v_l/R$ ($R=20\AA$).}
\end{figure}
\par
\emph{Anisotropic spin model and DM spin exchange interaction-} The general form of RKKY interaction in
Eq. (3) can divide into Heisenberg type $J^{i}$ and Dzyaloshinskii-Moriya
(DM) type $J^{DM}$,
\begin{equation}
H_{RKKY}=\sum_{i=x,y,z}J^{i}{S_{1}^{i}}{S_{2}^{i}}+J^{DM}\mathbf{e}\cdot
\left( \mathrm{\mathbf{S}_{1}}\times \mathrm{\mathbf{S}_{2}}\right)
\end{equation}%
where $\mathbf{e=}(\hat{x},\hat{y},\hat{z})$ is the unit vector and the
spin-frustrated terms $\sum_{i\neq
j=x,y,z}J^{fr}(S_{1}^{i}S_{2}^{j}+S_{1}^{j}S_{2}^{i})$ vanish if the
impurities are distributed along principle axes.
\par
The detailed spin textures for the RKKY interaction of S-DSMs, as well as DSMs, are shown in supplemental material\cite{appendix}. Compared to the case of DSMs, there exists a significant difference for the short-range (or weak $u_F$) behavior of the RKKY interaction, i.e., a stronger magnetic anisotropy would arise in S-DSMs in the condition of small $k_FR$ as impurities are deposited on the relativistic axis. Specifically speaking, the RKKY components $J^i_l$ are anisotropic with $J^x_l\neq J^z_l\neq J^y_l$, which would generate the $XYZ$ spin model for all S-DSMs and distinguishes itself from the $XXY$ (e.g., $J^x_l= J^z_l\neq J^y_l$) spin model of DSMs. The underlying physics is attributed to the competition of the RKKY interactions with different decaying rates. Taking 2D S-DSM as an example, the RKKY components of finite Fermi energy in the long-range limit exhibits an anisotropy of $J^{x}_l=J^z_l\neq J^{y}_l$ [see Eq. (11) of the supplemental material\cite{appendix}], where $J^{x,z}_l$ falls off as $R^{-3/2}$ and $J^y_l\propto R^{-2}$. When the short-range limits (small $R$) is considered, the higher-order terms $R^{-2}$ with different coefficients in $J^{x,z}_l$ become considerable, which compete with the term of $R^{-3/2}$ and lead to $J^x_l\neq J^z_l\neq J^y_l$.
\par
Next, we focus on the
influence of the anisotropic dispersion on the DM term. Starting from Eq. (4),
the DM coefficient can be obtained with
\begin{equation}
J^{DM}\mathbf{e}=-\frac{4\lambda ^{2}}{\pi }\mathrm{Re}\int_{-\infty
}^{u_{F}}d\omega G_{\mathrm{0}}(G_{x},G_{y},G_{z}).
\end{equation}
Here, the appearance of DM term must satisfy two conditions: (1) Finite
Fermi energy. If $u_{F}=0$, all the models show the vanished DM interaction
due to the protected electron-hole symmetry, similar to the case of WSMs\cite%
{Weyl1,Weyl2}. (2) Breaking the symmetry of spatial inversion. Note that $%
G\left( \mathbf{R},\omega \right) =\sum_{i=0,x,y,z}G_{i}\left( \mathbf{R}%
,\omega \right) \sigma _{i}$. If impurities are distributed along the
square-momentum direction, only diagonal components $G_{0}$ and $G_{z}$ in
the spin space are nonzero, i.e., $G\left( \mathbf{R},\omega \right)
=G_{0}\sigma _{0}+G_{z}\sigma _{z}$. It is easy to confirm $G\left( -\mathbf{%
R},\omega \right) =G\left( \mathbf{R},\omega \right) $ and so $J^{DM}=0$.
Once the impurities are deposited along the linear-momentum direction, the
nondiagonal components $G_{x}$ and $G_{y}$ are included, which breaks the
inversion symmetry, $G\left( -\mathbf{R},\omega \right) \neq G\left( \mathbf{%
R},\omega \right) $, and so finite DM interaction emerges. In Fig. 3, we plot the dependence of DM exchange interaction for 2D on impurity direction $\theta_R$. Compared with isotropic DM interaction for the DSMs, the DM interaction for S-DSMs exhibits strong anisotropic, which is largest for linear-momentum direction ($\theta_R=\pi/2$) and vanishes for $\mathbf{k}^2$ direction ($\theta_R=\protect\pi/2$). These results also are in agreement with derived analytical expressions for limit cases (see derivation in supplemental material\cite{appendix}). Also, the introduction of non-relativistic contribution in the S-DSMs can reduces the dependence of DM interaction on $u_{F}$, in comparison with the case of DSMs.
\par
If the Hamiltonian of Eq. (8) is expressed in the pseudospin space [$\sigma_i\rightarrow \tau_i$ in Eq. (8)], taking orbital space as an example, the term ${\rm Tr}[\sigma_\alpha\sigma_i\sigma_j\sigma_\beta]$ in Eq. (4) have to be rewritten as ${\rm Tr}[\sigma_\alpha\sigma_\beta]{\rm Tr}[\tau_i\tau_j]$ according to the Refs.\cite{Dirac1,Weyl1}. Thus, the DM terms would vanish and only the Heisenberg-type RKKY interaction survives with an isotropic $XXX$ ($J^x_{l(p)}=J^y_{l(p)}=J^z_{l(p)}$) spin model, similar to the case of graphene\cite%
{graphene1,graphene2,graphene3,graphene4,graphene5}. But the decay of the RKKY interaction still follows the law shown in Eqs. (10-13).
\par
\emph{Conclusions-} We have theoretically explored the
RKKY interaction between magnetic impurities in S-DSMs including all 2D and
3D models. Due to the coexistence of the relativistic and non-relativistic
electrons, the RKKY interaction of S-DSMs is anisotropic and violates the
decaying law proposed in isotropic systems. We find that the introduction of non-relativistic electrons in the S-DSMs can significantly prolong the decay of the RKKY interaction along relativistic direction, in comparison with the case of isotropic DSMs. For example, the decaying rate $R^{-5}$ for 3D DSMs is reduced to be $R^{-3}$ for type-II 3D S-DSMs, which can greatly facilitate the experiment detection and magnetic doping technology. The underlying physics is ascribed
to the interplay of relativistic and non-relativistic electrons.
Furthermore, we give a general formula to determine the decaying rate from
the system dimension and the non-relativistic dimension. In addition, we
find that the anisotropy of S-DSMs can greatly affect the DM component of the RKKY interaction, which is largest for impurities in the relativistic direction but vanishes in non-relativistic direction.
\par
All these peculiar magnetic characteristics implies that the S-DSMs are a powerful platform to detect and control the magnetic exchange interaction, superior to extensively adopted isotropic systems. Experimentally, a variety of candidates for S-DSMs with
different approaches have been proposed, such as multilayer $(\mathrm{%
TiO_{2}})_{m}/(\mathrm{VO_{2}})_{n}$ nanostructures\cite%
{Victor1,Victor2,Banerjee}, deformed graphene\cite{press2}, and silicene oxide\cite%
{oxide}. The RKKY interactions can be probed experimentally with present techniques, e.g., spin-polarized scanning tunneling spectroscopy\cite{Laplane}, which
can measure the magnetization curves of individual atoms, or
the electron-spin-resonance technique coupled with an optical
detection scheme\cite{Wiebe1,Wiebe2}.

\acknowledgements This work was supported by GDUPS (2017), by the National
Natural Science Foundation of China (Grants No. 12047521, Grants No.
11874016 and No. 11904107), by the Science
and Technology Program of Guangzhou (No. 2019050001), by the Guangdong NSF of China (Grants No. 2021A1515011566 and No. 2020A1515011566), by Guangdong Basic and Applied
Basic Research Foundation (Grant No. 2020A1515111035), and by the Project
funded by China Postdoctoral Science Foundation (Grant No. 2020M672666).

\end{document}